\documentstyle[aps,prl]{revtex}
\begin{document}
\twocolumn
\input{epsf}
\input psfig.sty
\draft
\title{Strong Confinement and Oscillations in Two-Component
Bose-Einstein
Condensates}
\author{Q-Han Park$^{1,2}$ and J. H. Eberly$^1$}
\address{$^1$ Rochester Theory Center for Optical Science and
Engineering
and \\
Department of Physics and Astronomy,\\
University of Rochester, Rochester, New York 14627-0171}
\address{$^{2}$ Department of Physics, Kyunghee University, Seoul,
130-701, Korea}

\date{\today}
\maketitle
\begin{abstract}
Abstract: We present a new model of BEC dynamics based on strong
confinement near the ground state. The model predicts oscillations in a
two-component condensate, based on interference of non-spreading wave
packets moving within a pair of tilted nearly square potentials. The
oscillations are similar to those recently reported for a magnetically
trapped $^{87}$Rb condensate, and the model's predictions give good
quantitative agreement with the experiments.
\end{abstract}

\vspace{.25 in}
Since its first application to quantum Rabi oscillations in the
two-level cavity QED context, collapse and revival
analysis \cite{ebenarsan} has been used for the understanding of a
widening class of recurrent physical phenomena.
The timing and nature of revivals have been employed in interpreting
the dynamics of a variety of quantum systems \cite{various1,various2},
but a number of cases provide puzzling features
that are considered open questions \cite{puzzles}.

Recent elegant experimental observations of Rabi-type oscillations in a
two-component Bose-Einstein condensate of $^{87}$Rb \cite{Matthews}
offer a new example that has resisted analysis. Multi-component BEC's
\cite{JILA,MIT}
in the presence of external driving fields have provided a rich spectrum
of physical responses to challenge current theoretical understanding. In
the two-component case of interest here \cite{Hall1998a,Hall1998b} the
theoretical situation has been described by various groups
\cite{Hall1998b,Dum1998a}.
In particular, differential shifting of the trap centers for two species
has led to interesting theoretical predictions such as a ``condensate
dressed state" \cite{Blakie}, Josephson-type oscillations and
periodic modulation of Rabi oscillations \cite{Williams}, all of which
originate in analogies with two-level quantum optical systems. Despite
heuristic arguments using ``twists" of an $SU(2)$ order parameter
\cite{Matthews}, it is fair to say that the fundamental dynamics of the
observed oscillations are not understood.

In this Letter, we go beyond revival analysis to present a new quantum
mechanical picture of the physics based on a strong confinement model,
which explains experimentally observed features quite well. We show that
the coupled Gross-Pitaevski (GP) equations for the two-component
Bose-Einstein condensate can be decoupled by a time-dependent dressing
transformation. The transformed Schr\"{o}dinger equations describe localized
wavepackets that accelerate in opposite directions and bounce back and
interfere after hitting the walls of the confinement. We find that the
spatial and temporal behavior of the two BEC's are determined by these
wavepackets and the experimental observations that have been reported can
be explained in detail in terms of their interference. Our model gives
quantitative predictions that also agree well with numerical simulation
of the coupled GP equations.

The two components of the condensate, $\psi_1$ and $\psi_2$, obey GP
equations that are coupled to each other via the ``drive" parameter
$\Omega$, which plays a role similar to the Rabi frequency of a driven
two-level system:
\begin{equation}
i \frac{\partial \psi_1}{\partial t} = H_0 \psi_{1}
+ \mu (r_{1} |\psi_{1}|^2 + |\psi_{2}|^2)\psi_{1}
+ {\delta \over 2}\psi_{1} + {\Omega \over 2}\psi_{2}
\end{equation}
\begin{equation}
i{\partial \psi_{2} \over \partial t}
= H_0\psi_{2} +\mu (|\psi_{1}|^2 +r_{2} |\psi_{2}|^2 )\psi_{2} -{\delta
\over 2}\psi_{2} + {\Omega \over 2}\psi_{1},
\label{GP}
\end{equation}
where $H_0 \equiv -{1\over 2}\frac{\partial^2 }
{\partial z^2 } + {1\over 2}z^2 $ is the bare Hamiltonian for the
harmonic trap. For simplicity we present here only a one dimensional model.
We have used the axial oscillation frequency $\omega_{z}$ to express units:
$\hbar \omega_{z}$ is our unit of energy and $1/\omega_{z}$ and $d_z
=\sqrt{\hbar/ m\omega_{z}}$ are the units of time and distance. The
coefficient $\mu = 4\pi a_{12}N/d_z$ measures the strength of
interaction between two species of BEC where $N$ is the total particle number
and $a_{12}$ is the interspecies scattering length. The small difference reported
\cite{Matthews} for the ratios in ${}^{87}Rb$, $r_{1} =a_{1}/a_{12}
\approx 1.03$ and $r_{2} =a_{2}/a_{12} \approx 0.97$, among scattering
lengths has little effect on the subsequent discussion so we assume that
$r_{1}=r_{2}=1$ in the following.

The detuning $\delta$ and the interspecies coupling $\Omega$, both
measured in the unit $\omega_{z}$, can depend on the axial position $z$.
We will make assumptions of no detuning ($\delta =0$) and a linear
$z$-dependence of coupling: $ \Omega = \Omega_{0} + 2\beta z $,
consistent with experiment \cite{Matthews}.  The nonvanishing detuning
case will be discussed later. Now we introduce the time-dependent
dressing transformation
\begin{equation}
\phi_{\pm} \equiv (\psi_{1} \pm \psi_{2} )e^{\pm i
\Omega_{0}t/2},
\end{equation}
which gives two decoupled GP equations in a slow time frame:
\begin{equation}
i {\partial \phi_{\pm } \over \partial t}
= -{1\over 2}{\partial ^2 \phi_{\pm } \over \partial z^2 }
+ V_{\pm }\phi_{\pm} ,
\label{phipm}
\end{equation}
where the potentials $V_{\pm}$ are given by
\begin{equation}
V_{\pm }(z) = {1\over 2} z^2  + {\mu \over 2}
( |\phi_{+}|^2 + |\phi_{-}|^2 ) \pm \beta z .
\end{equation}
The first crucial observation is that although these potentials
are spatially very nonlinear, they are nearly independent of time if the
system is started near its ground state. This is due to the strong and
relatively tight nonharmonic confinement provided in the present case by
the GP interaction. In principle, the potentials depend on time through
the total density of the condensate, $\rho_{tot} = |\psi_{1}|^2+|\psi_{2}|^2
= {1\over 2}(|\phi_{+}|^2+|\phi_{-}|^2) $, but $\rho_{tot}$ changes very
little during hundreds of Rabi cycles that are well-resolved in the
experiments. This is evident in the relatively very rigid spatial shape
of the total condensate density seen in the series of snapshots in Fig. 1.
To obtain the snapshots we solved the coupled GP equations using the
combined split-step Fourier technique familiar from fiber optics
theory (where the GP equation is called the non-linear Schr\"odinger
equation), modified to include two components \cite{GPA}

The second crucial observation, also evident in Fig. 1 and noted previously
\cite{Hall1998a}, is that $\rho_{tot}$
maintains a Thomas-Fermi-like spatial distribution. We have found that
this allows the confining forces on the packets to be converted into
boundary conditions, leaving relatively free $\phi_{\pm}$ packets to accelerate
slowly in opposite directions at the rate $\pm \beta$ along the bottoms of two
potential wells that are nearly square but tipped in opposite
directions, as shown in Fig. 2. We will now demonstrate that this simplified
picture is powerfully effective in unraveling relatively complicated condensate
evolution.

If, as in the JILA experiments, a condensate at nearly zero temperature
is prepared initially in the pure 1 component, i.e. $\psi_{1}(z,0)=\psi_g(z),~
\psi_{2}(z,0) = 0$, we have equally distributed $\phi's: ~\phi_{+}(z,0)
= \phi_{-}(z,0) = \psi_g(z)$. As these $\pm$ probability packets fall
along the bottom of the well toward opposite walls they must interfere since
$\phi_{+}$ and $\phi_{-}$ are coherent parts of the same two-component
condensate. An effective interaction between the decoupled $\pm$ packets
comes from their interference to produce the observed density
$|\psi_1|^2$:
\begin{eqnarray}
|\psi_{1}|^2 &=& {1 \over 4}( |\phi_+|^2+|\phi_-|^2 )
+ {1\over 2}
\Re\left(e^{i\Omega_{0}t}\phi_{+}^{*}\phi_{-}\right)\nonumber \\
&\approx& {1 \over 2} |\psi_g|^2
+ {1\over 2} \Re\left(e^{i\Omega_{0}t} \phi_{+}^{*}\phi_{-}\right),
\label{den1}
\end{eqnarray}
where $ \Re$ denotes the real part.

First we show additional results of direct numerical integration of the
original two nonlinear GP equations (\ref{phipm}). The calculated
temporal evolution of the space-integrated population of component 1 is
shown in Fig. 3. The corresponding experimental data is given in Fig. 4-a.
Rapid oscillations at the drive
frequency $\Omega_0$ are evident, as the population cycles between
components 1 and 2. An apparent collapse in region $A$, and a revival or
perhaps a pair of them, is seen in regions $D$ and $E$. Small features
at $B$ and $C$ are also noted.  Now we will use the strong confinement
model to obtain a detailed understanding of the curve. As we will see,
revivals in the conventional sense play no role. The correct explanation
is in some respects simpler, and capable of wider applicability as we mention at
the end.

Now we return to our simplified picture, according to which the main
slow motion is acceleration with each packet moving toward an opposite wall.
This main slow motion can also be factored out of the quantum amplitudes by
defining
\begin{equation}
\chi_{\pm}(z,t) = \phi_{\pm}(z,t) \exp( \pm i\beta z t +i \beta^2 t^3
/6),
\end{equation}
and changing the coordinates into those of the two accelerating frames:
$T=t, ~ Z_{\pm} =z \pm \beta t^2 /2$. The packets $\chi_{\pm }$ satisfy
the free Schr\"{o}dinger equation, $\left(i{\partial / \partial T} + {1
\over
2}{\partial^2 / \partial Z_{\pm}^2 }\right) \chi_{\pm } = 0$, with
initial conditions $\chi_{\pm}(T=0) = \psi_{g}$, and so can be integrated
directly to yield
\begin{equation}
\chi_{\pm}={1 \over 2\pi }\int \int ~dz'dk ~e^{ik(z-z' \pm \beta t^2 /2)
- i k^2 t/2} \psi_g(z',0) .
\end{equation}
Since the initial $\psi_g$ is very smooth, this shows that the two
$\chi_{\pm}$ move apart with a combined
quantum mechanical dispersion which is relatively small in our
Thomas-Fermi like configuration. The interference term in Eq.(\ref{den1}) is now
given by ${1\over 2}\Re\left(\exp[i\Omega_{0}t+2i \beta z t]
\chi_{+}^{*}\chi_{-}\right)$ with $\chi_{\pm}$ as given above.

Thus we predict spatial oscillations with wavelength $L_{0} = \pi /\beta
t$ to develop in the population of condensate 1, dramatically evident in
Fig. 1-B. The interference occurs in the region where the $\pm$ packets
continue to overlap, a shrinking region of length $L = 2R-\beta t^2$, where
the full axial length is $2R$. These spatial
interference oscillations have nodes that can be monitored
experimentally [see Fig. 4-b] and the node formation has been called
``twisting." Various effects are predicted and labeled in Fig. 3 as follows.

A. {\it Collapse.} The number of nodes in existence at time $t$ is fixed
at
$n = L/L_{0} = \beta t(2R-\beta t^2)/\pi $. By about the time of the
second twist there is a good balance of the two components in the
condensate (e.g., $|\psi_1|^2 \approx |\psi_2|^2 \approx 0.5$), which
shows up in the data as the completion of the ``collapse" at the end
of region A in Fig. 3. From $n=2$, with the
parameters used in the calculation ($R=8.53, ~\beta =0.25$), one easily
finds for the collapse time $t_C \approx 1.5$, in excellent agreement
with Fig. 3.

B. {\it Twist resolvability.} Our formula also predicts a maximum  twist
number $n_M$, and says that it occurs at time $t_{M} = \sqrt{2R/3\beta}$,
after which the $\pm$ packets are falling too fast to have resolvable
spatial interference. This gives $n_{M}= \sqrt{32 \beta R^3 /27 \pi^2}$,
or $t_{M}=4.77$ and $n_{M}=4.3$, also in good agreement with the observed
behavior in region B.

C. {\it Bounce time, $t=t_B$.} When the centers of $\phi_{\pm }$ reach
the walls at time $t_{B}=\sqrt{2 R/\beta}$ (${} = 8.26 $ in our case), the
$\phi_{\pm}$ become maximally separated and have a finite but minimally
overlapping interference region. Eq. (\ref{den1}) shows that in the
near-wall regions, the densities $|\psi_{1}|^2$ and $|\psi_{2}|^2$
reduce to about half of the initial ground state density. This cannot be seen
in Fig. 3, but is confirmed by examining snapshots of the spatial behavior
of the two condensate components. Five such snapshots were already given in
Fig. 1, and Fig. 1-C confirms that at the time of the wall bounce, in
the regions near the walls, the two components are not interfering but each
contain about half of the total density. Of course, due to reflection at
the wall and subsequent self-interference, there arise complicated
motions for each $\phi_{\pm}$. A more complete analytic description of the
interference between $\phi_{\pm}$ can be made by solving the
Schr\"{o}dinger equation (\ref{phipm}) directly in terms of Airy
functions.

D. {\it Packet recovery, $t=t_R$.} After bouncing from the strong
confinement, the packets $\phi_{\pm}$ return approximately to their
initial positions. As they return, oscillations in the overlapping region
spread out over the half-ground-state density region with increasing amplitude
and wavelength. When the number of nodes reaches three, with one central
peak and two side peaks, oscillations in the half-ground-state region
interfere most destructively, thereby minimizing the population of
component 1  (Fig. 1-D). This happens before completing one round trip
by the amount of time required to develop three nodes in the initial stage.
However, the external coupling drive quickly switches back from the
minimum population to the maximum population for condensate 1. Thus,
the maximal recovery occurs at $t_{R}=2t_{B} -\Delta t$ where $\Delta t$ is
obtained from $\beta \Delta t (2R-\beta (\Delta t )^2 )/\pi =3$.

In our case, $\Delta t=2.4$ and $t_{R}=14.1$. This shows that the
maximum revival of population does not come with the original shape. In fact,
Fig. 1-D shows that it rather has the shape of a dark soliton. This is also
in agreement with the reduced contrast of about 60 percent of the original.

E. {\it Completing one cycle, $t=t_{F}$.} At the
classical return time, $t_{F}=2t_{B}$, the recovery of population of
condensate 1 happens but in a smaller scale than in case $D$, as
shown in Fig. 1-E.

In closing, let us return to an earlier point. In the spin-boson
language of revival analysis, a vanishing detuning ($\delta = 0$ in the
evolution equation (\ref{GP}), for example) corresponds to exactly zero Zeeman
splitting of the two spin states, removing any possibility of resonant
coincidence with the boson (oscillator) transition frequencies. This
is why revival anaysis is misapplied to the existing data. However,
although the BEC oscillator is very anharmonic under its strong
confinement, one suspects that it could support a near-resonance with a
non-zero $\delta$, if $\delta$ were large enough. In such a case one
would have an intriguing combination of bouncing packets and quantum revivals
(which themselves are of course known to have packet interpretations).
Since the energy spacings in the confinement are small, $\delta$ would
not have to be very large.

In this Letter, we have introduced a quantum mechanical packet model
which provides an excellent qualitative description of the previously
unexplained temporal behavior of oscillations in two-component driven-BEC
experiments. One clearly sees that the strong-confinement bouncing packet model
permits easy analysis that predicts all five distinct events in the temporal
record as well as predicting their appropriate time scales. Figs. 1-3 compare
these events with numerical simulations and with the experimental record
in Fig. 4.

It is clear that this model is not restricted to the two-component case,
and it will be interesting to extend our work to the three-component BEC
recently reported \cite{MIT} and partially analysed \cite{Pu}.

We thank S. Raghavan for his participation in the early stage of this
work and C.R. Stroud, Jr. for a key remark about Airy packets. This
research is supported by NSF grant PHY94-15583. Q.P. is also
supported in part by the Brain Korea 21 Project, and by KOSEF 97-07-02-02-01-3.

\begin{figure}
\vglue .1in
\leftline{\epsfxsize 1.6in\epsfclipon\epsffile{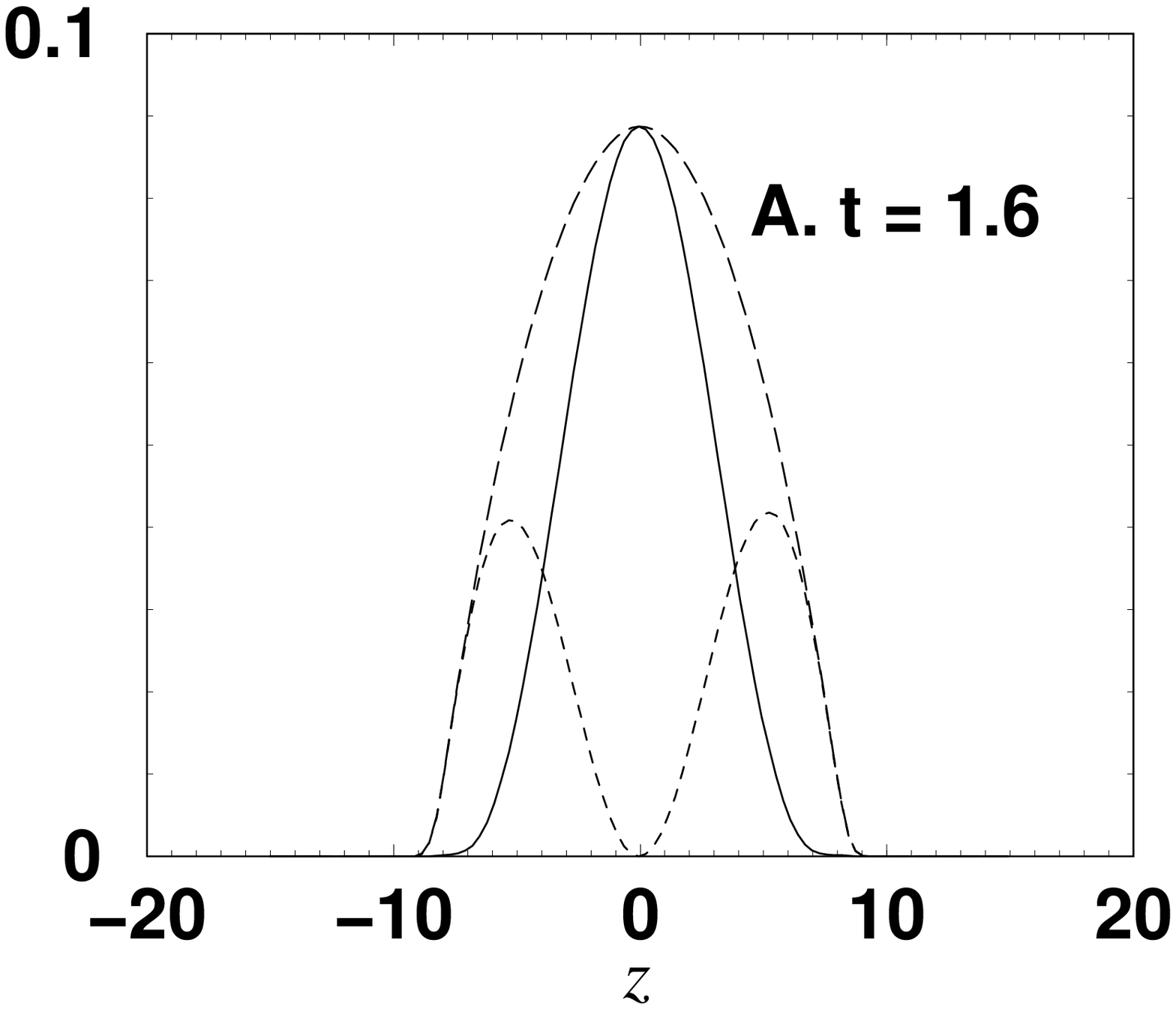}}
\vglue -1.4in
\rightline{\epsfxsize 1.6 in \epsfclipon\epsffile {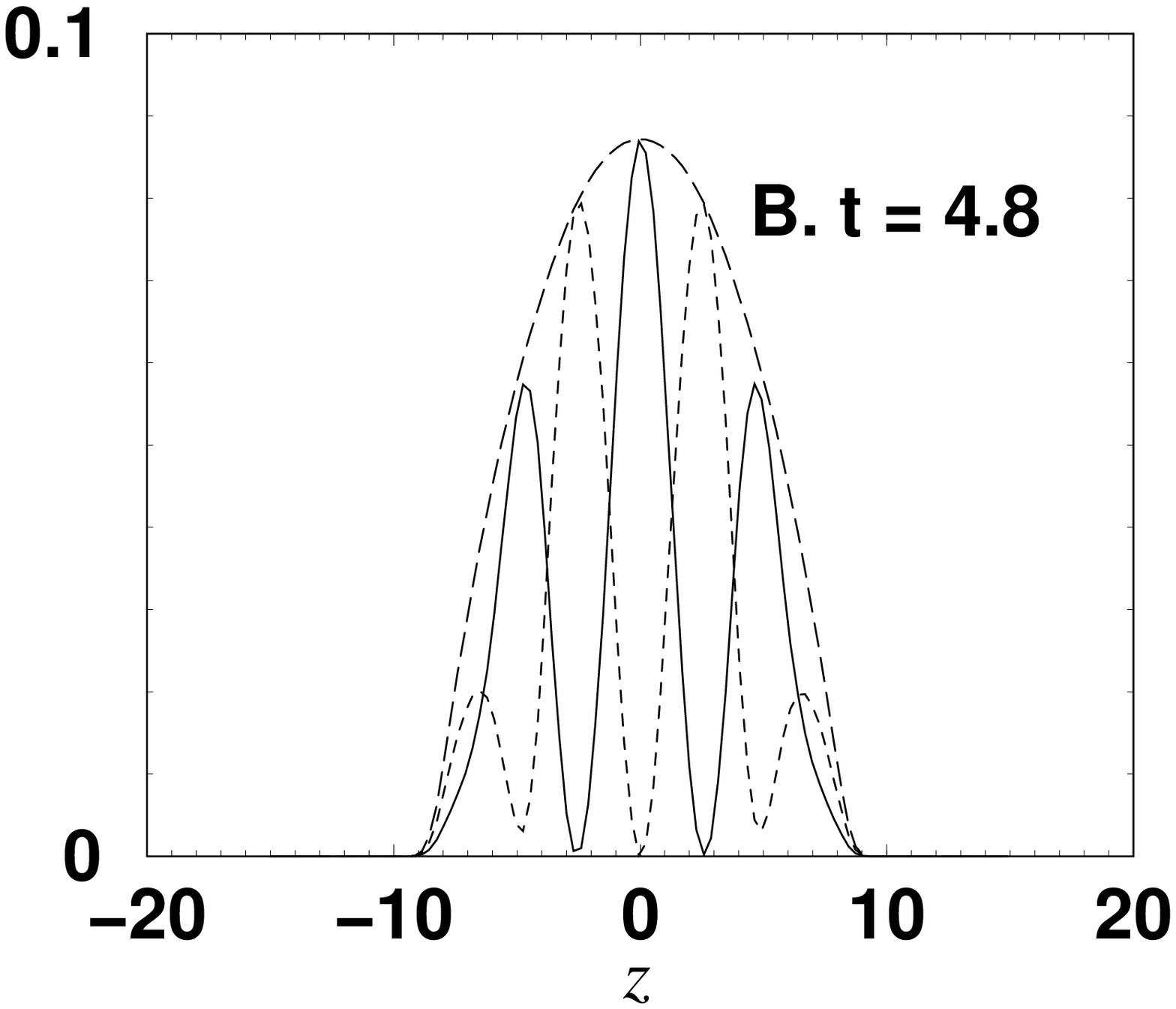}}
\vglue .3 in
\leftline{\epsfxsize 1.6 in\epsfclipon\epsffile {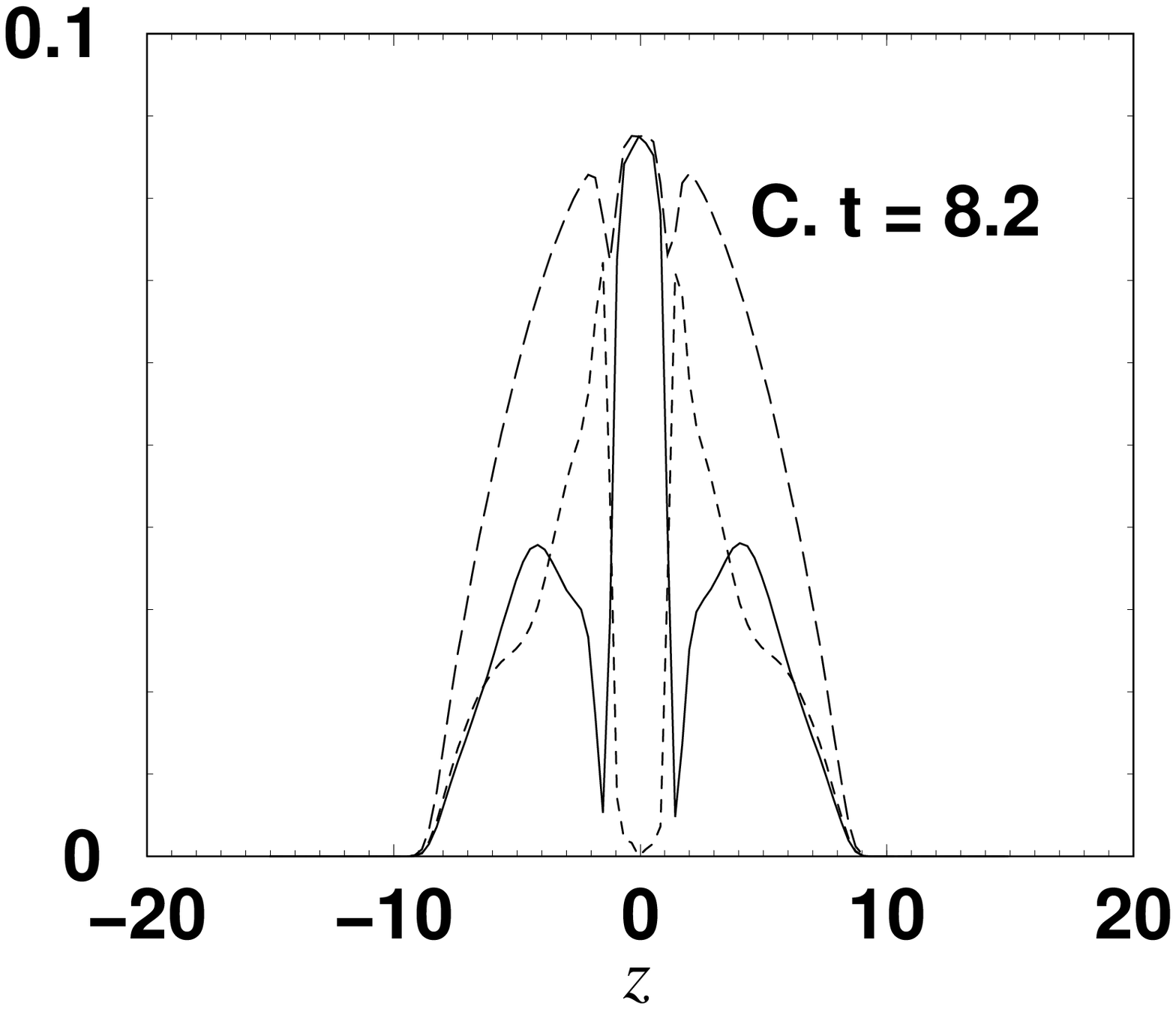}}
\vglue -1.4 in
\rightline{\epsfxsize 1.6 in\epsfclipon\epsffile {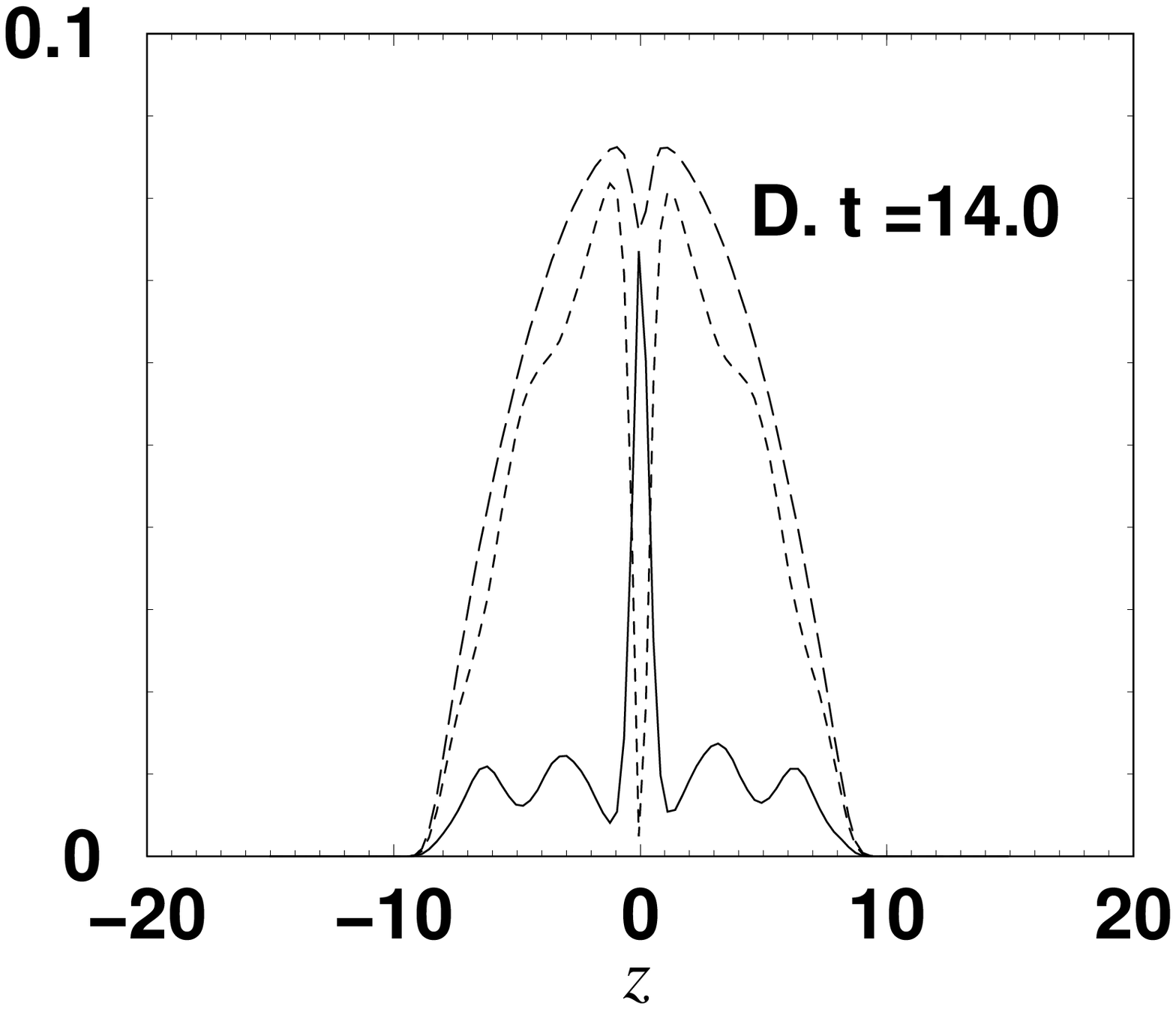}}
\vglue .3 in
\leftline{\epsfxsize 1.6 in \epsfclipon\epsffile {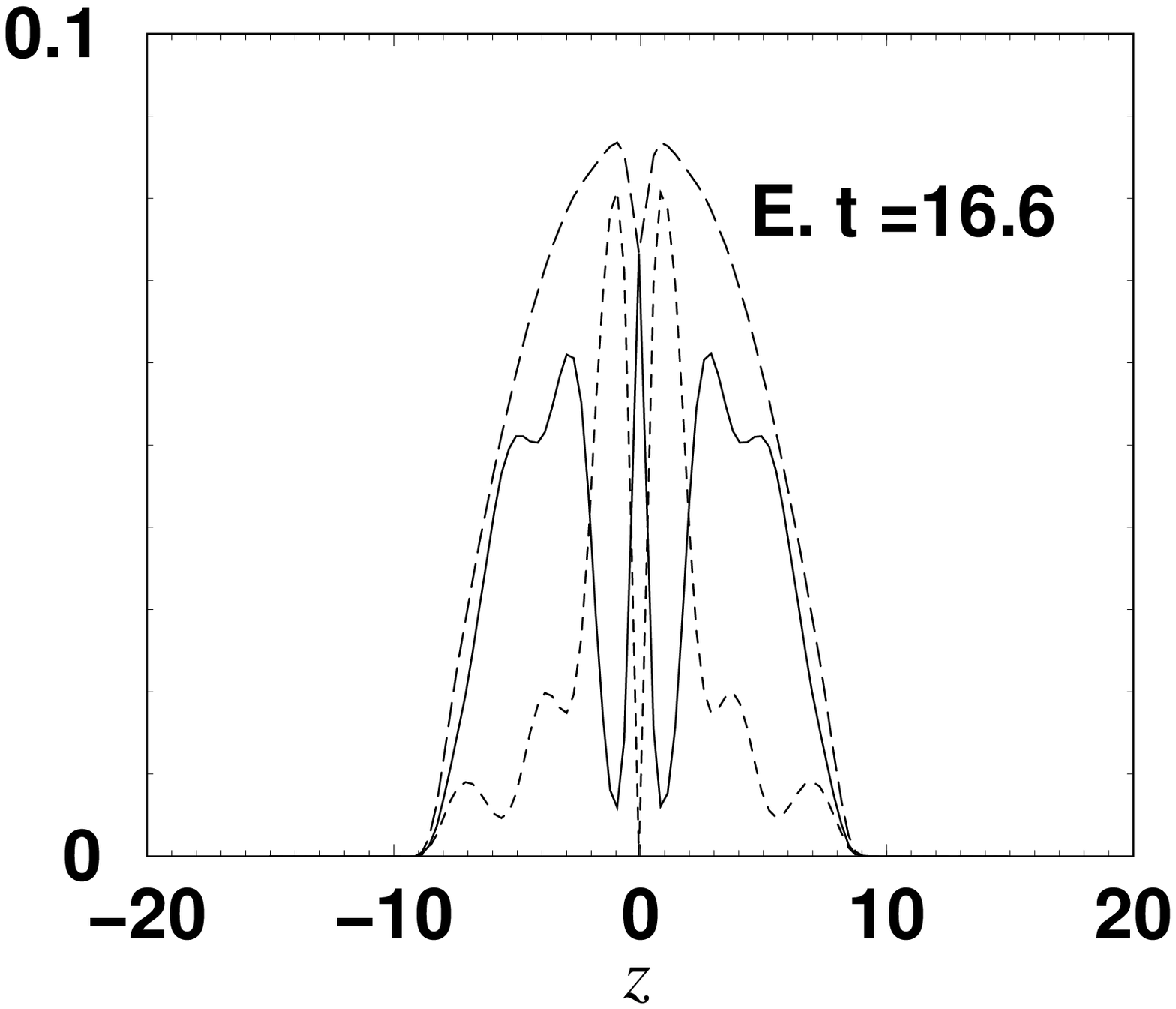}}
\vglue .3in
\caption{Calculated density profiles of condensates at five distinct
times that are the keynote instants during evolution noted in Fig. 3.
Solid and dotted lines represent condensates 1
and 2, and the dashed line represents the total density $\rho_{tot}$.
Note the strong time-independence of $\rho_{tot}$.}
\end{figure}
\vglue .2in
\begin{figure} \vglue .1in
\centerline{\epsfxsize 2.2 in \epsfclipon\epsfbox {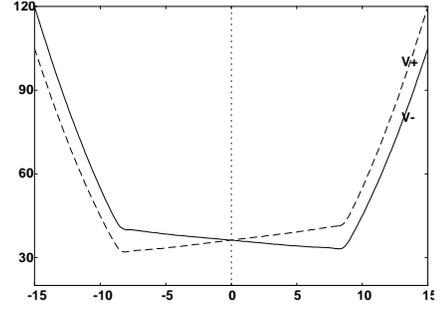}} \vglue .5in
\caption{Potentials $V_{\pm}$. Since the total
density maintains the Thomas-Fermi shape, the potentials have a flat
bottom while the wall is the upper portion of a harmonic potential. }
\end{figure}

\begin{figure}[p]
\begin{center}
\psfig{figure=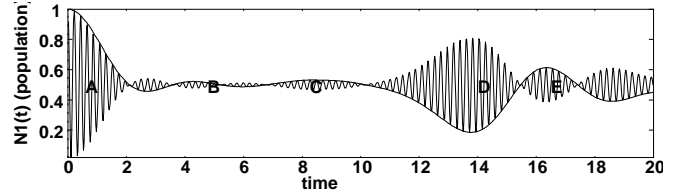,width=1\linewidth,clip=}
\vglue .2in
\caption{Plot of the population of condensate 1. The thick line
represents the case with no constant coupling drive ($\Omega_{0}=0$)
while the fast oscillating thin line represents the case with
$\Omega_{0}=28.85$.
Five distinct events are specified at the calculated times, which are
described separately in the text.}
\end{center}

\end{figure}

\vglue .2in
\begin{figure}[p]
\begin{center}
\psfig{figure=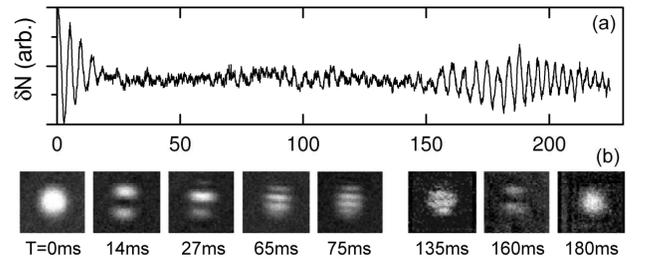,width=1\linewidth,clip=}
\end{center}
\caption{Experimental data from ref. [5]. The modest discrepancy in time
registry copared to Fig. 3 is due to our simplified 1D modeling and also to
an experimental departure from $\delta = 0$, evident because the mid-line
in part 4(a) shows a component 1 fraction lower than 50\%.}
\end{figure}
\end{document}